\documentclass[11pt]{article}

\usepackage[final]{acl}

\usepackage{times}
\usepackage{latexsym}

\usepackage[T1]{fontenc}
\usepackage[utf8]{inputenc}

\usepackage{microtype}

\usepackage{inconsolata}
\usepackage[table]{xcolor}
\usepackage{booktabs}

\usepackage{graphicx}

\title{Evaluating Multi-turn Human-AI Interaction}

\author{
Shi Ding\thanks{Accepted at the ACL 2026 Workshop on Evaluating Evaluations (EvalEval).} \\
Georgia Institute of Technology \\
Atlanta, Georgia, USA \\
\texttt{sding84@gatech.edu}
\And
Sijian Tan \\
Georgia Institute of Technology \\
Atlanta, Georgia, USA \\
\texttt{stan43@gatech.edu}
}

\begin{document}
\maketitle
\begin{abstract}

Large language models (LLMs) are increasingly used as collaborative assistants, yet dominant NLP evaluation practices remain centered on aggregate metrics such as accuracy and fluency. These approaches often overlook behaviors that are critical in human-facing settings (e.g., consistency across multiple turns and iterative refinement). In this paper, we examine limitations of current NLP evaluation practices and introduce TCR, a structured framework for evaluating human--AI interaction using educational LLM assistants as an illustrative example. TCR emphasizes dimensions such as transparency, consistency, and refinement. We further present structured evaluation prompts and illustrative interaction examples demonstrating how structured evaluation can complement aggregate metrics and LLM-as-a-judge approaches. Our work highlights the need for more human-centered evaluation practices for interactive LLM systems.
\end{abstract}

\section{Introduction}
Large language models (LLMs) are increasingly adopted in human-facing settings, yet their evaluation remains largely governed by general-purpose NLP benchmarks. While these benchmarks support scalable and standardized comparison, they often provide limited insight into interactive behaviors that emerge in real-world use\cite{ding2026bridging}. This paper examines gaps in current LLM evaluation practices and introduces \textbf{TCR (Transparency, Consistency, and Refinement)}, a structured framework for analyzing model behaviors in multi-turn human--AI interaction. Using educational settings as an illustrative example, we discuss how structured evaluation may reveal limitations often overlooked by dominant NLP evaluation approaches and outline implications for the design and assessment of interactive LLM systems. We take the position that aggregate-centric evaluation may overlook critical behaviors in multi-turn interaction, highlighting the need for more interpretable behavioral evaluation. We make two contributions in this position paper. First, we identify gaps between dominant NLP evaluation practices and real-world interactive AI systems. Second, we introduce TCR, a structured framework that supports interaction-level evaluation beyond aggregate correctness metrics for interactive LLM systems.

\section{Current Model Evaluation}
\subsection{Aggregate Evaluation}

Dominant NLP evaluation practices have traditionally emphasized correctness-oriented metrics such as accuracy, precision, recall, and F-score measures \cite{hutchinson2022evaluation}. These approaches primarily rely on reference-based metrics, such as BLEU \cite{papineni2002bleu} and ROUGE \cite{lin2004rouge}, as well as reference-free approaches including LLM-as-a-judge frameworks. While these methods support scalable and standardized comparison, they remain limited for evaluating open-ended and human-facing generation tasks\cite{kashidlarge}. 

LLM-as-a-judge frameworks have recently gained prominence due to their scalability\cite{wang2018glue,wang2019superglue}. However, prior work shows that LLM judges exhibit systematic biases that can distort evaluation outcomes \cite{li2025generation,szymanski2025limitations}. In many cases, these approaches continue to prioritize surface-level fluency and preference signals while providing limited insight into how interaction quality evolves across multi-turn and human-facing contexts. Benchmarks such as MT-Bench \cite{zheng2023judging} and Arena-Hard-Auto \cite{li2024crowdsourced} exemplify this trend by emphasizing conversational performance while providing limited insight into model behavior across real-world interaction contexts.

While single-turn evaluation has long been dominant, recent work shows that static benchmarks can obscure demographic biases and failures in AI-mediated interaction\cite{star2026evaluating}, Recent work has increasingly explored multi-turn and long-horizon evaluation settings beyond static single-turn benchmarks. but often remain centered on preference modeling, or aggregate outcome optimization rather than broader human-centered interaction qualities \cite{chen2026intermt,abdulhai2026consistently}. Prior work further emphasizes that meaningful evaluation should account for broader contextual and human-centered considerations beyond benchmark performance \cite{selbst2019fairness,mitchell2019model,raji2021ai}. Together, these considerations highlight that meaningful evaluation should account not only for performance outcomes, but also for the broader assumptions, behaviors, and consequences that emerge when language models are deployed in human-facing settings.

\subsection{Human-Facing Evaluation}

Prior research argues that evaluating machine learning systems requires attention not only to model outputs, but also to human behavior, institutional practices, and deployment contexts \cite{ji2023ai,hendrycks2020aligning,shen2024towards}. Interactive benchmarks such as Chatbot Arena \cite{chiang2024chatbot} introduce more dynamic evaluation settings, but often remain outcome-centric and rely heavily on aggregate success metrics. Human-facing LLM systems frequently require contextual guidance, iterative interaction, and adaptation across multiple turns. However, dominant evaluation approaches still emphasize correctness and fluency metrics, providing limited insight into interaction quality and evolving user context. Educational settings provide an illustrative example of these limitations because effective instructional support depends on iterative interaction, contextual guidance, and learner-centered communication \cite{vygotsky1978mind,kapur2008productive,ding2025considering}. Correctness alone provides limited insight into whether AI systems support understanding, refinement, or productive interaction over time \cite{shneiderman2022human,friedman2002value}.
Recent work explores multi-turn tutoring evaluation and personalized interaction modeling in educational AI. However, these approaches often remain centered on aggregate optimization and reward-based evaluation rather than structured analysis of interaction quality \cite{abdulhai2026consistently,wan2026enhancing}. Educational settings therefore illustrate why human-facing multi-turn interaction may require more interpretable and process-oriented evaluation approaches.

Existing evaluation approaches for human-facing LLM systems are often either aggregate-centric or too coarse to capture model behaviors in authentic multi-turn settings. Unlike prior work that evaluates isolated usability or explainability factors, TCR focuses specifically on interaction-level behaviors emerging across multi-turn human--AI interaction. This position paper builds on a prior scoping review examining how AI agents are evaluated in human-facing environments \cite{levac2010scoping}. The review analyzed 126 sources across HCI, AI, and education venues and identified recurring evaluation themes\cite{ding2025rethinking}. Rather than proposing a broad evaluation taxonomy, this paper focuses on interaction quality in human-facing multi-turn LLM systems using education as an illustrative example.

\section{TCR Evaluation Framework}
\subsection{Framework Overview}

We introduce TCR as a lightweight framework for evaluating model behaviors in multi-turn human-facing AI systems. Rather than focusing only on traditional aggregate correctness or fluency, TCR emphasizes how models support users through iterative interaction, contextual guidance, and refinement across turns. The framework focuses on behaviors that are often overlooked in dominant NLP evaluation practices, particularly in settings where users rely on AI systems for explanation, clarification, and ongoing support. TCR also includes structured evaluation prompts designed to support more interpretable and process-oriented assessment practices. As shown in Table~\ref{tab:evaluation-comparison}, TCR emphasizes dynamic multi-turn behaviors that are often overlooked by traditional evaluation approaches.

% In document
\begin{table}[htbp]
\small
\centering
\renewcommand{\arraystretch}{1.25}

\begin{tabular}{m{3.2cm}m{3.2cm}}
\toprule

\cellcolor{gray!12}\textbf{Traditional Evaluation}
&
\textbf{TCR}
\\

\midrule

\cellcolor{gray!8}Correctness
&
{\color{blue!50!black}\textbf{Transparency}}
\\[4pt]

\cellcolor{gray!8}Preference
&
{\color{green!40!black}\textbf{Consistency}}
\\[4pt]

\cellcolor{gray!8}Task Completion
&
{\color{orange!70!black}\textbf{Refinement}}
\\

\bottomrule
\end{tabular}

\caption{Comparison between traditional evaluation approaches and the dynamic multi-turn behaviors emphasized by TCR.}
\label{tab:evaluation-comparison}

\end{table}

\subsubsection{Evaluation Dimensions}

TCR focuses on three behavioral dimensions that commonly emerge in human-facing multi-turn interaction: transparency, consistency, and refinement. These dimensions are presented as an intentionally scoped framework rather than an exhaustive taxonomy, while remaining extensible across additional domains and interaction contexts.

\textbf{Transparency}
Transparency refers to whether an agent communicates reasoning in clear and interpretable ways \cite{nauta2023anecdotal, arrieta2020explainable, guidotti2019stability}. In educational settings, explanations are part of the learning process rather than auxiliary outputs. Relevant indicators include reasoning clarity, traceability, and alignment between explanations and task outcomes \cite{ silva2023explainable, bommasani2021opportunities, bransford2000people, amershi2019guidelines, radanliev2025privacy,kim2026transparency}.

\textbf{Consistency}
Consistency refers to the stability of system behavior across similar conditions \cite{nauta2023anecdotal, nielsen1995heuristics}. While benchmarks often report aggregate scores, they may overlook behavioral variance across prompts, runs, or evaluators. Relevant indicators include output stability and sensitivity to prompt perturbations\cite{carvalho2019machine, robnik2018perturbation, vanden2012increasing,lu2025simplifying,shen2025boosting}.

\textbf{Refinement}
Refinement refers to an agent’s ability to support iterative interaction through clarification, revision, and feedback over time \cite{wang2024ali, pan2024autonomous}. Unlike static single-turn evaluation, educational interaction is inherently iterative. Relevant indicators include responsiveness to feedback and coherence across revisions\cite{guo2024using, wang2024ali, hong2024my, nam2026mle,cai2025automated}.

While these dimensions are defined conceptually above, Table~\ref{tab:framework-indicators} summarizes illustrative indicators for how they may appear in practice. The framework focuses on educational contexts as an example of human-facing multi-turn interaction.

\begin{table}[htbp]
\small
\centering
\renewcommand{\arraystretch}{1.25}

\begin{tabular}{p{2.3cm}p{4.8cm}}
\toprule
\textbf{Dimension} & \textbf{Example Indicators} \\
\midrule

\cellcolor{blue!8}\textbf{T} Transparency
& Step-by-step reasoning; source references \\[4pt]

\cellcolor{green!8}\textbf{C} Consistency
& Output stability under prompt paraphrases; variance across runs \\[4pt]

\cellcolor{orange!10}\textbf{R} Refinement
& Responsiveness to feedback; clarity of revision rationale \\

\bottomrule
\end{tabular}

\caption{Illustrative indicators associated with TCR dimensions.}
\label{tab:framework-indicators}
\end{table}

While TCR is presented conceptually in this work, the framework may also support structured evaluation through lightweight annotation practices. These dimensions may further enable hybrid human--AI evaluation workflows and analyses such as agreement comparison across evaluators.

%While TCR is presented conceptually in this work, it may also be interpreted as a structured behavioral evaluation formulation for human-facing LLM systems:

%\[
%f(c,x) = (t,s,r)
%\]

%where \(c\) represents dialogue context, \(x\) represents the model response, and \(t\), \(s\), and \(r\) denote transparency, consistency, and refinement scores, respectively, with \(t,s,r \in \{1,2,3\}\). Here, 1 indicates low alignment, 2 indicates partial alignment, and 3 indicates strong alignment. This formulation illustrates how TCR dimensions may support structured hybrid evaluation and standard statistical analyses such as agreement analysis.

\subsubsection{Structured Evaluation Prompts}

TCR may also support reflective evaluation through lightweight structured prompts \cite{jacovi2020towards, shneiderman2022human}. These prompts illustrate how evaluators may move beyond outcome-focused judgments toward more process-oriented assessment. Table~\ref{tab:guiding-questions} presents example prompts for each dimension.

\begin{table}[htbp]
\small
\centering
\renewcommand{\arraystretch}{1.2}
\setlength{\tabcolsep}{5pt}

\begin{tabular}{p{2.1cm}p{4.8cm}}
\toprule
\textbf{Dimension} & \textbf{Example Prompts} \\
\midrule

\cellcolor{blue!6}\textbf{Transparency}
& Does the AI explain its reasoning in a clear and understandable way? \\[-1pt]
& Are reasoning steps or supporting information visible to the user? \\[5pt]

\cellcolor{green!6}\textbf{Consistency}
& Does the system remain consistent across similar prompts or interaction contexts? \\[-1pt]
& Are outputs stable across sessions or evaluators? \\[5pt]

\cellcolor{orange!8}\textbf{Refinement}
& Does the AI support iterative revision and improvement over multiple turns? \\[-1pt]
& Does the refinement process remain aligned with the user’s goals and context? \\

\bottomrule
\end{tabular}

\caption{Illustrative reflective prompts associated with TCR dimensions.}
\label{tab:guiding-questions}
\end{table}

\subsubsection{Evaluation Example}

To illustrate behaviors that may be overlooked by aggregate evaluation, Table~\ref{tab:interaction-example} presents an example multi-turn interaction with an LLM programming assistant. While both responses may appear technically correct under aggregate or LLM-as-a-judge evaluation, the example highlights differences in transparency, contextual continuity, and refinement-oriented support across turns.

\begin{table}[htbp]
\footnotesize
\centering
\renewcommand{\arraystretch}{1.2}
\setlength{\tabcolsep}{6pt}

\begin{tabular}{p{1cm}p{6.3cm}}
\toprule
\textbf{Turn} & \textbf{Interaction} \\
\midrule

1 &
\textbf{User:} Why does my \texttt{fitMedia} code not play any sound? \\[4pt]

& \cellcolor{orange!10}
\textbf{Model A:} This issue may be related to how the audio variable is referenced in your program. In programming, variables need to remain consistent across function calls and media operations. \\[4pt]

& \cellcolor{blue!8}
\textbf{Model B:} Check whether the sound variable name inside \texttt{fitMedia()} matches the variable you created earlier. For example, if you created \texttt{AK\_UNDOG\_808\_1}, the same name should also appear inside \texttt{fitMedia()}. \\[6pt]

\midrule

2 &
\textbf{User:} I do not understand how to fix it. \\[4pt]

& \cellcolor{orange!10}
\textbf{Model A:} Variable consistency is important because programming functions depend on correct parameter references and object assignment. \\[4pt]

& \cellcolor{blue!8}
\textbf{Model B:} First, find the line where you created the sound variable. Then check the name used inside \texttt{fitMedia()}. The two names should match exactly. Even one missing letter in \texttt{AK\_UNDOG\_808\_1} can stop the sound from playing. \\[6pt]

\midrule

\multicolumn{2}{p{7.3cm}}{
\cellcolor{gray!10}
\textbf{Evaluation Summary:}
Aggregate or LLM-as-a-judge evaluation may rate both responses as generally correct, while TCR highlights differences in how the models support novice users through concrete guidance, step-by-step clarification, and iterative interaction.
} \\[4pt]

\bottomrule
\end{tabular}

\caption{Illustrative multi-turn interaction and evaluation comparison.}
\label{tab:interaction-example}
\end{table}

\section{Implications for Model Evaluation}

Our analysis highlights key limitations in dominant model evaluation practices, particularly the common assumptions around correctness, fluency, and automated judgment in the evaluation of human-facing LLM systems. We distill two implications that clarify why current evaluation practices remain insufficient for interactive language models.

\subsection{Implication 1: Limitations in LLM-as-a-Judge Systems}

LLM-as-a-judge frameworks offer scalability and efficiency for automated evaluation. However, our analysis identifies limitations in their applicability to human-facing evaluation settings. Evaluating interaction quality often requires sensitivity to context, user goals, and iterative interaction patterns that cannot be reliably inferred through fluency- and similarity-based evaluation alone. In our illustrative example, LLM judges may prioritize technically correct responses while overlooking contextual misalignment, inconsistent behavior, or ineffective refinement support. These patterns suggest that LLM-as-a-Judge Systems provide only a partial view of interaction quality and highlight the need for hybrid evaluation approaches that combine automated auditing with structured human-centered assessment.

\subsection{Implication 2: Evaluating Multi-Turn Interaction}

Current evaluation approaches often collapse multi-turn interaction into aggregate correctness or preference signals, masking behaviors that emerge through refinement, clarification, and contextual adaptation over time. While these approaches support scalable comparison, they provide limited insight into how interactions evolve through dialogue, revision, and iterative refinement. These limitations highlight the sociotechnical nature of evaluation, which is shaped by human values, institutional expectations, and contextual interaction. As illustrated in our educational example, structure evaluation through structured evaluation prompts helps surface assumptions that are often hidden by aggregate evaluation practices and makes model behavior more transparent in interactive contexts.

Together, these implications suggest that future model evaluation methods should incorporate contextual evaluation, multi-turn consistency analysis, and hybrid human--AI assessment approaches.

\section{Limitations and Future Work}

While TCR is intentionally scoped in this work, future work may further validate and extend the framework through large-scale empirical studies and more accessible evaluation practices for diverse users and interaction contexts.


\bibliography{custom}

\end{document}